T = 37 °C

$(G_{min} - \mu)/\sigma$

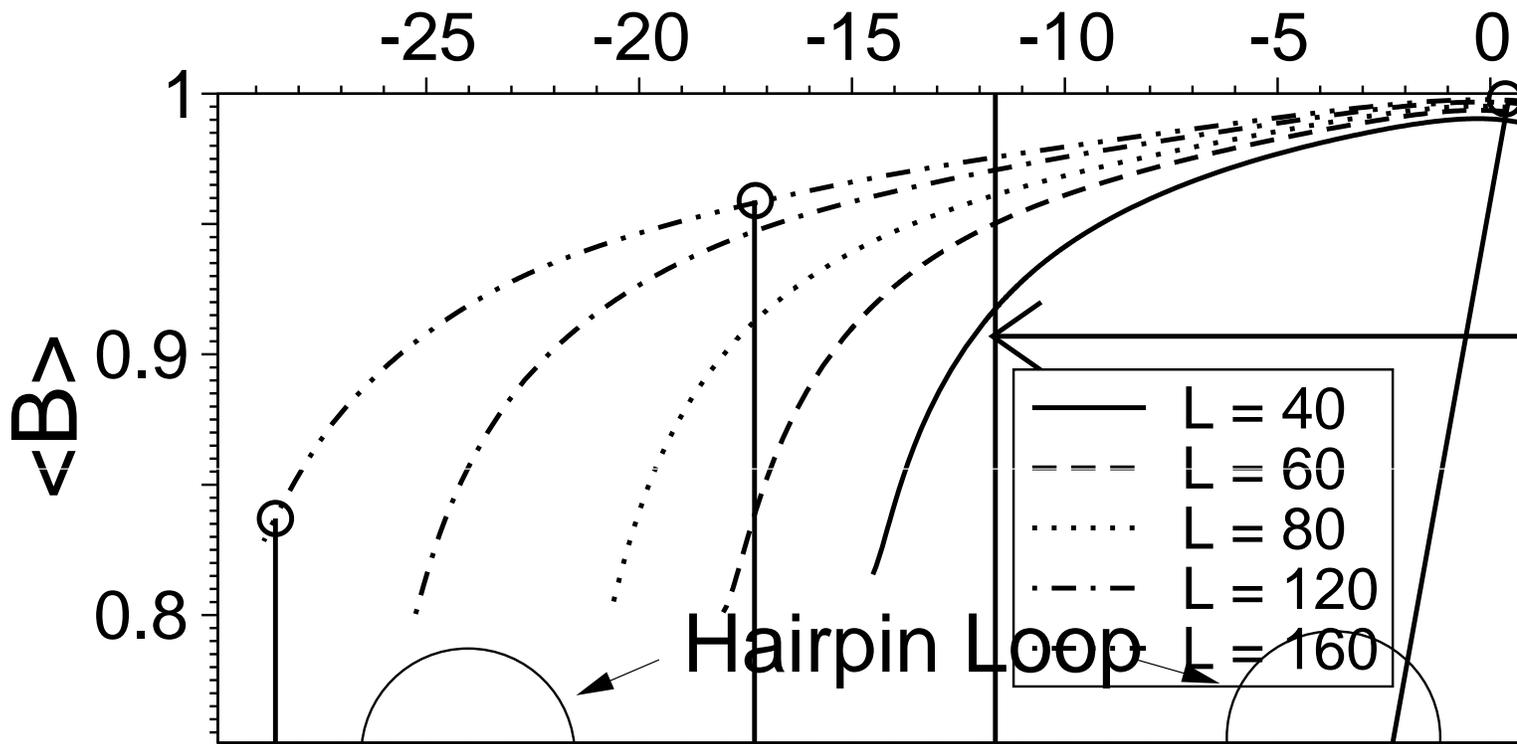

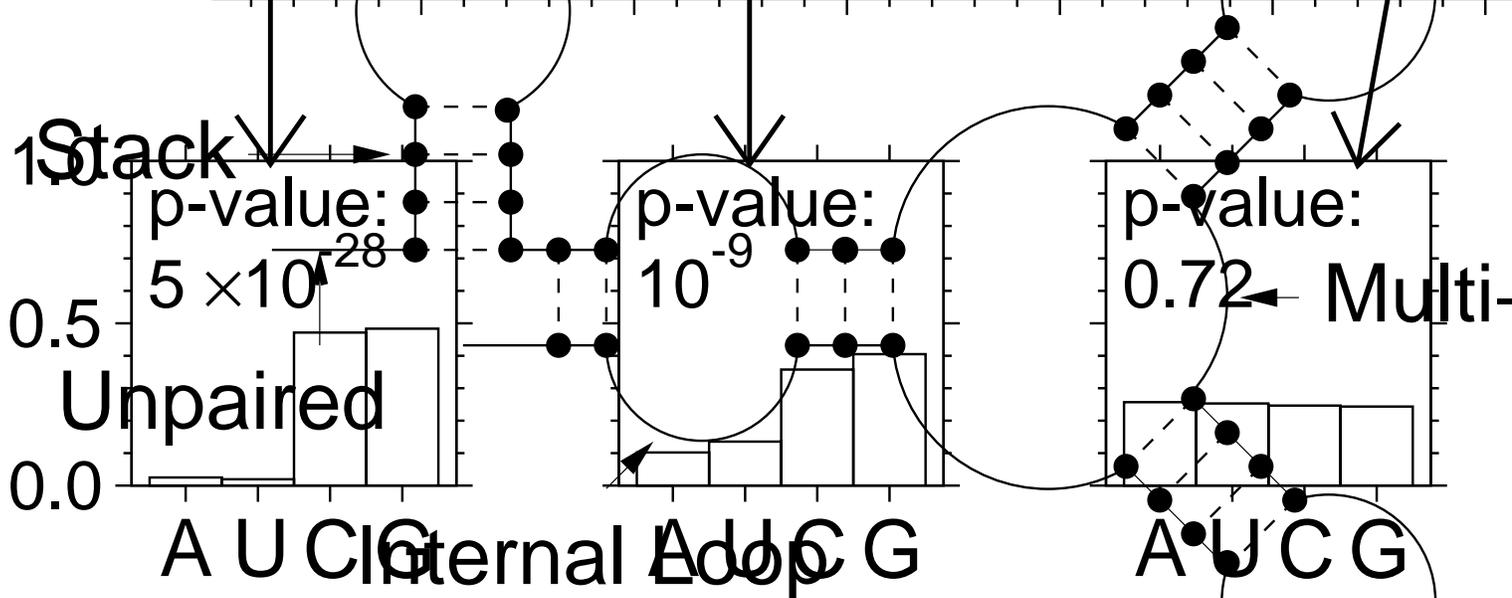

Stack
Unpaired

Internal Loop

Multi-

p-value: $5 \times 10^{-28}$

p-value: $10^{-9}$

p-value: 0.72

Hairpin Loop

L = 40
L = 60
L = 80
L = 120
L = 160

$\langle B \rangle$

A U C G    A U C G    A U C G

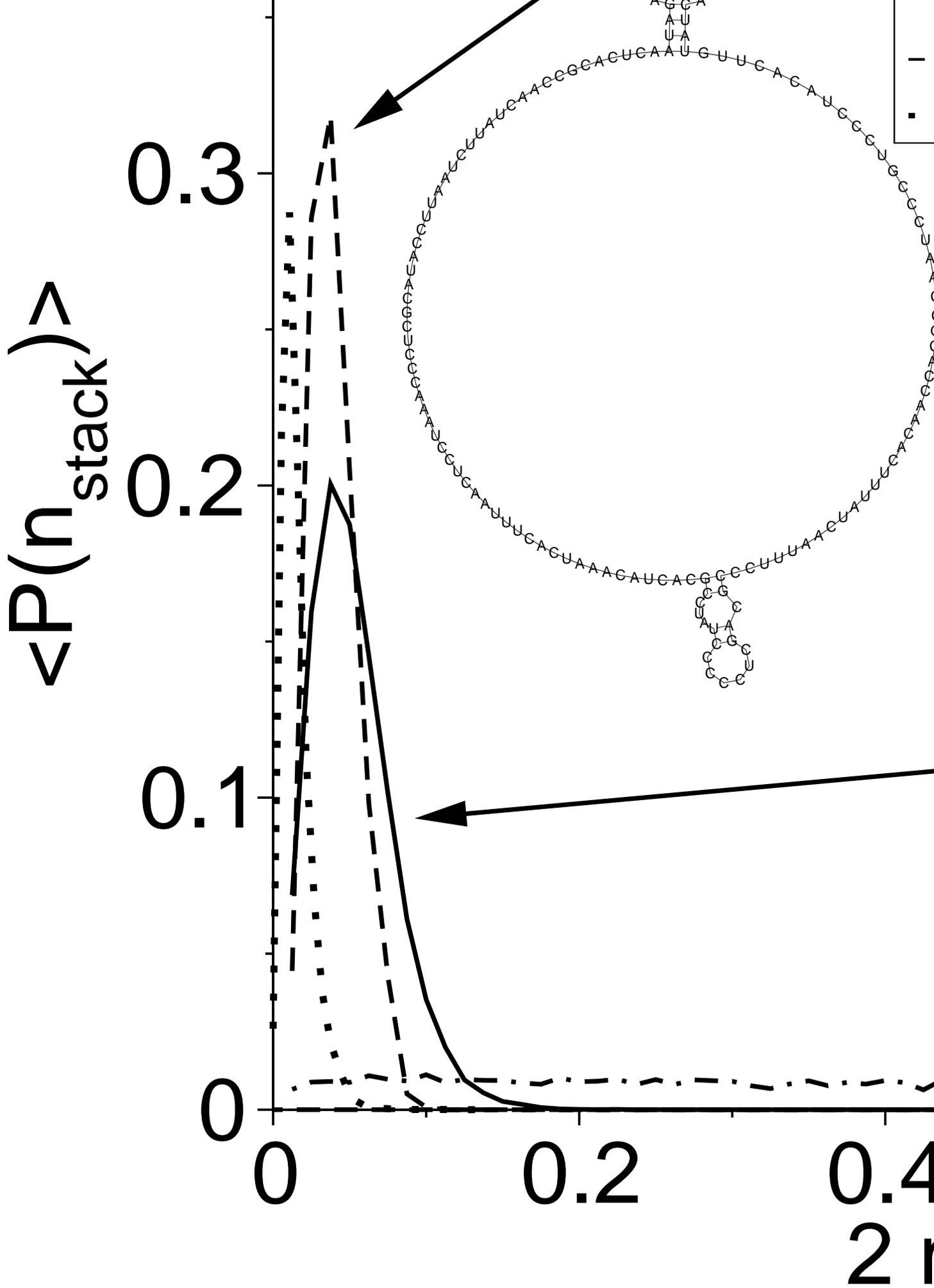